\documentclass{pasa}

\title[Unexpected Circular Radio Objects]{Unexpected Circular Radio Objects at High Galactic Latitude}

\author[Norris et al.]{
Ray P. Norris$^{1,2\star}$, 
Huib T. Intema$^{3,4}$, 
Anna D. Kapi\'{n}ska$^{5}$, 
B\"arbel S. Koribalski$^{1,2}$, 
Emil Lenc$^{2}$, 
L. Rudnick$^{6}$, 
Rami Z. E. Alsaberi$^{1}$, 
Craig Anderson$^{5}$,
G. E. Anderson$^{3}$,
E. Crawford$^{1}$,
Roland Crocker$^{7}$,
\alter{Jayanne English$^8$,}
Miroslav D. Filipovi\'c$^{1}$,
\alter{Tim J. Galvin$^3$,}
Andrew M. Hopkins$^{9}$,
Natasha Hurley-Walker$^{3}$,
Susumu Inoue$^{10}$,
Kieran Luken$^{1,2}$,
Peter J. Macgregor$^{1,2}$, 
Pero Manojlovi\'c$^{1}$,
Josh Marvil$^{5}$,
Andrew N. O'Brien$^{1,2,11}$,
\revb{Laurence Park}$^1$,
Wasim Raja$^{2}$,
Devika Shobhana$^{1}$,
\alter{Tiziana Venturi$^{12}$,}
Jordan D. Collier$^{1,13}$,
Catherine Hale$^{2}$,
Aidan Hotan$^2$,
Vanessa Moss$^{2}$,
\& Matthew Whiting$^2$
 \thanks{ e-mail: raypnorris@gmail.com }

 \affil{$^1$ Western Sydney University, Locked Bag 1797, Penrith, NSW 2751, Australia}
 \affil{$^2$ CSIRO Astronomy \& Space Science, P.O. Box 76, Epping, NSW 1710, Australia}
 \affil{$^3$ International Centre for Radio Astronomy Research, Curtin University, GPO Box U1987, Perth, WA 6845, Australia}
 \affil{$^4$ Leiden Observatory, Leiden University, PO Box 9513, NL-2300RA, Leiden, The Netherlands}
 \affil{$^5$ National Radio Astronomy Observatory, PO Box 0, Socorro, NM87801, USA}
 \affil{$^6$ University of Minnesota, 100 Church St SE, Minneapolis, MN 55455, USA}
\affil{$^7$ Australian National University, Canberra ACT 2600, Australia}
\affil{$^8$ University of Manitoba, Winnipeg, Manitoba Canada, R3T 2N2.}
 \affil{$^9$ Australian Astronomical Optics, Macquarie University, 105 Delhi Rd, North Ryde, NSW 2113, Australia}
 \affil{$^{10}$ iTHEMS, RIKEN, 2-1 Hirosawa, Wako, Saitama, 351-0198, Japan}
 \affil{$^{11}$ Center for Gravitation, Cosmology and Astrophysics, Department of Physics, University of Wisconsin-Milwaukee, PO Box 413, Milwaukee, WI 53201, USA}
 \affil{$^{12}$ Istituto di Radioastronomia, INAF, Via Gobetti 101, 40129 Bologna, Italia}
 \affil{$^{13}$ The Inter-University Institute for Data Intensive Astronomy (IDIA), Department of Astronomy, University of Cape Town, Private Bag X3, Rondebosch, 7701, South Africa}

}

\jid{PASA}
\doi{10.1017/pas.\the\year.xxx}
\jyear{\the\year}

\usepackage{aas_macros}
\usepackage{hyperref} 
\hypersetup{colorlinks,citecolor=blue,linkcolor=blue,urlcolor=blue}

\usepackage[utf8]{inputenc} 
\usepackage[left]{lineno} 
\usepackage{cite} 
\usepackage{graphicx} 
\usepackage{url} 
\usepackage{lscape} 
\usepackage{xcolor}

\newif\iffinal
\iffinal\renewcommand{\includegraphics}[2][]{}\fi

\newcommand\nar{{New Astron. Rev.}}

\newcommand{\kms}{\,km\,s$^{-1}$} 

\newcommand{\rev}[1]{{#1}}
\newcommand{\revb}[1]{{#1}}
\newcommand{\alter}[1]{{#1}}

\newcommand{\lsun}{\ifmmode{{\rm ~L}_\odot}\else{~L$_\odot$}\fi}
\newcommand{\Msun}{\ifmmode{{\rm ~M}_\odot}\else{~M$_\odot$}\fi}
\newcommand{\degr}{$^{\circ} $}
\newcommand{\sqdeg}{\,deg$^2$}

\newcommand{\ujybm}{$\mu$Jy/beam}

\begin{document}
\begin{frontmatter}
\maketitle

\begin{abstract}
We have found a \alter{class of circular radio objects}
in the Evolutionary Map of the Universe Pilot Survey, using the Australian Square Kilometre Array Pathfinder telescope. 
The objects appear in  radio images as circular edge-brightened discs, about one arcmin diameter, \alter{that are unlike other objects previously reported in the literature. We explore several possible mechanisms that might cause these objects, but none seems to be a \revb{compelling}  explanation. 
}


\end{abstract}

\begin{keywords}
Extragalactic radio sources 
\end{keywords}
\end{frontmatter}

\setlength{\baselineskip}{15pt}

\section{Introduction}

Circular features are well-known in radio astronomical images, and usually represent a spherical object such as a supernova remnant, a planetary nebula, a circumstellar shell, or a face-on disc such as a protoplanetary disc or a star-forming galaxy. They may also arise from imaging artefacts around bright sources caused by calibration errors or inadequate deconvolution.  Here we report the discovery of a class of circular feature in radio images that do not seem to correspond to any of these known types of object or artefact.
For brevity, and lacking an explanation for their origins, we dub these objects ``Odd Radio Circles'', or ORCs.

These objects were first discovered in the Pilot Survey   \citep{norris20} of the  Evolutionary Map of the Universe   \citep[EMU:][]{emu}, which is an all-sky continuum survey using the newly-completed Australian Square Kilometre Array Pathfinder telescope (ASKAP)  \citep{johnston07,johnston08,mcconnell16}.  

Three ORCs (ORCs 1--3) were discovered by visual inspection of the images from the survey. Their rarity, together with their low surface brightness, makes it unlikely that they could have been discovered in previous radio surveys.  

We discovered a further ORC (ORC~4) 
in archival data taken with the Giant MetreWave Radio Telescope (GMRT)  \citep{gmrt} in March 2013. 
In most respects it is very similar to ORCs 1--3, but differs in having a central radio continuum source. 

We \rev{observed three} of the ORCs with the Australia Telescope Compact
\revb{Array \citep[ATCA: ][]{wilson11} and the Murchison Widefield Array    \citep[MWA:][]{tingay13}
to provide independent confirmation of the reality of these objects, and  to provide spectral index information.}

The EMU Pilot Survey area is covered by Data Release~1 of the Dark Energy Survey   \citep[DES:][]{abbott18} and we use DES data throughout this paper, together with infrared  data from the Wide-field Infrared Survey Explorer (WISE)  \citep{wise}, and, for ORC~4, optical data from the Sloan Digital Sky Survey SDSS  \citep{sdss}. 

\alter{In Sections 2 and 3 of this paper we describe the observations and results. In Section 4 we present the observational properties of the ORCs, and in Section 5 we discuss possible causes of this phenomenon.}

\alter{Throughout this paper we assume a flat $\Lambda$CDM cosmology, based on \citet{planck18}, with the following parameters: 
 H$_0$ = 67.36.
$\Omega_m$ = 0.3153,
$\Omega_\Lambda$ = 0.6847.} For a source with flux density S at frequency $\nu$, we define spectral index $\alpha$  as $S \propto \nu^\alpha$.

\section{Observations}

\subsection{ASKAP Observations}
\label{obs}
ASKAP consists of $36 \times 12$-m antennas, each of which is equipped with a chequerboard Phased Array Feed (PAF), giving a field of view of about 30 square degrees of the sky, resulting in a high survey speed.  A full description of ASKAP may be found in  Hotan et al. (2020; in preparation).

The EMU Pilot Survey   \citep{norris20} (hereafter EMU-PS) consists of ten observations, each lasting 10--12 hours, in the period 15 July to 24 November 2019. Between 32 and 36 ASKAP antennas were used in each observation, always including the outer four, with baselines up to 6.4 km, with the remaining antennas within a region of 2.3 km diameter.  We used 36 PAF beams, with the beam centres separated by 0.9 degrees, arranged in the ``Closepack36'' configuration   \citep{norris20}. Specifications of the observations are given in Table~\ref{specs}.

\begin{table}
\begin{center}
\caption{EMU Pilot Survey Specifications }
\begin{tabular}{ll}
\hline
Area of survey          & 270 \sqdeg  \\
Field centre            & 21h, --55 degr \\
Synthesised beamwidth   & $13'' \times 11''$ FWHM \\
Frequency range         & 800 -- 1088 MHz \\
Observing configuration & closepack36, pitch 0.9\degr \\
 & \mbox{no interleave} \\
Weighting               & Robust = 0 \\
RMS sensitivity         & 25 -- 35 \ujybm \\
Total integration time  & 100 hours \\
Number of sources detected & $\sim$250,000 \\
\hline
\end{tabular}
\label{specs}
\medskip\\
\end{center}
\end{table}

We processed the data using the ASKAPsoft pipeline   \citep{norris20, whiting17}, including w-projection multifrequency synthesis imaging, multiscale clean, and self-calibration.

\subsection{ATCA Observations}

We observed ORC~1 and ORCs~2--3 (Project code C3350) with the Australia Telescope Compact Array (ATCA) on 9--10 April 2020, at 1.1--3.1 GHz (weighted central frequency after the removal of radio frequency interference = 2121 MHz), over a period of 2 $\times$ 12 hours using the 6A configuration.  The data were processed using miriad, using the standard ATCA multi-frequency synthesis process, and cleaned with a robustness of +0.5. The observations were affected by radio frequency interference,  resulting in  a relatively high rms of $\sim$ 10--15 \ujybm, with a synthesised beamsize of 5.0 $\times$ 4.3 arcsec.

\subsection{GMRT Observations}
ORC-4 was found in archival 325 MHz GMRT data taken on the cluster source project (ID 23 017)   \citep{venturi17}  on the cluster Abell 2142. The data were reprocessed  with the SPAM pipeline   \citep{intema17}  yielding a sensitivity of 47 \ujybm\  at the field center, and a resolution of 9.4 $\times$ 7.9 arcsec. At the location of the ORC, the sensitivity is 66 \ujybm, mainly due to primary beam attenuation. 

\subsection {MWA Observations}
The observations with the Murchison Widefield Array   \citep{tingay13} in its `Phase II' extended configuration   \citep[MWA-2:][]{wtt+18} were taken as part of project G0045 which aims to image diffuse, non-thermal radio emission in galaxy clusters across five frequency bands of $\Delta\nu = 30$~MHz centered on 88, 118, 154, 185, and 216 MHz. As these observations have a large field of view, ORC1 and ORC2 fall within the primary beam main lobe half power point in one of the observed fields: `FIELD4'. MWA observations of this form are taken in a 2-min snapshot mode due to a fixed primary beam. All data undergo radio frequency interference flagging using \texttt{AOFlagger}  \citep{offringa15}. To increase integration time on a source, large numbers of 2-min snapshots are independently calibrated with the full-Jones \texttt{Mitchcal} algorithm  \citep{offringa16}, independently imaged, and finally stacked in the image plane as a linear mosaic.

MWA-2 data reduction for this work follows the process described in detail by \citet{duchesne2020}, using a purpose-written MWA-2 pipeline (\texttt{piip} \footnote{\url{https://gitlab.com/Sunmish/piip}}).

\subsection{Measurement of Flux Density and Spectral Index}
To measure flux densities, we used the measure\_source.py tool\footnote{\url{https://github.com/nhurleywalker/polygon-flux}}, in which a polygon is drawn around the source   \citep{2019PASA...36...48H}. \alter{Estimated uncertainties include a noise term and also the uncertainty in the flux density scale for each measurement.}

Spectral indices were derived by weighted least-squares fitting to the flux densities listed in the Tables, assuming a power-law spectral energy distribution. The uncertainty on the spectral index was estimated using the leastsq algorithm in scipy.optimize\footnote{\url{ https://docs.scipy.org/doc/scipy/reference/optimize.html}}.

\subsection{Optical/IR properties}

Photometry for the galaxies that may be associated with the ORCs were taken from the DES   \citep{abbott18},
WISE   \citep{wise}, 
and GALEX   \citep{galex}
surveys and are listed in Tables \ref{tab:orc1-wise} to \ref{tab:orc4-wise}.

\alter{The photometric redshifts listed in Tables \ref{tab:orc1-wise} to \ref{tab:orc4-wise} typically have quoted uncertainties of $\sigma_z \sim$ 0.01--0.02, but on the other hand we have found that independent photometric redshifts by different authors of radio sources in the EMU-PS field typically differ by 0.1--0.2, and so we regard this as being a more realistic uncertainty. }

\begin{table*}
\centering
\tiny
\caption{Properties of the optical/IR sources near ORCs 1--2 }.
\setlength{\tabcolsep}{3pt} 
\renewcommand{\arraystretch}{1.25} 
\label{tab:orc1-wise}
\begin{tabular}{cccccccccccccccl}

\hline
 & & ASKAP & \multicolumn{2}{c}{GALEX} & \multicolumn{5}{c}{DES} & \multicolumn{3}{c}{WISE} & \\
Source Name  & ID & flux & FUV & NUV & g & r & i & z & Y & W1 & W2 & W3 & W1--W2 & z & Notes\\
 & & [mJy] & \multicolumn{2}{c}{[mag]} & \multicolumn{5}{c}{[mag]} & \multicolumn{3}{c}{[mag]} & & \\
\hline
WISE J210258.15--620014.4 & ORC~1~C & $<$ 0.1 & --- & --- & 22.04 & 20.10 & 19.23 & 18.79 & 18.70 & 15.065  & 14.984  & $>$12.939 & 0.081 & \revb{0.551$^3$} & \\
 & & & & & 0.06 & 0.01 & 0.01 & 0.02 & 0.04 & $\pm$0.031 & $\pm$0.061 & & & &\\
WISE J210257.88--620046.3 & ORC~1~S & 0.86 & 23.7 & 22.3 & 19.733 & 18.945 & 18.550 & 18.351 & 18.311 & 15.472  & 15.063  & 11.201  & 0.409 & \revb{0.228$^3$}& \\
 & & & $\pm$1.2 & $\pm$0.3& 0.005 & 0.003 & 0.005 & 0.008 & 0.023 & $\pm$0.034 & $\pm$0.057 & $\pm$0.138 & &  & \\
WISE J205851.65--573554.1 & ORC~2~A & 1.0 & 25.9 & 20.9 & 17.676 & 17.355 & 17.263 & 17.253 & 17.352 & 16.038 & 16.501  & $>$12.716  & --0.463 &1.37$^2$ & listed as a star in Gaia DR2  \citep{gaia} \\ 
  & & & 4.0 & 0.1 & 0.001 & 0.001 & 0.001 & 0.002 & 0.01 & $\pm$0.050 & $\pm$0.252 & & &\\
WISE J205848.80--573612.1 & ORC~2~B & 1.7 & --- & --- & 20.53 & 19.03 & 18.52 & 18.20 & 18.09 & 15.138 &  14.995 & $>$12.431 & 0.143 & \alter{0.311$^1$} & 2.5 arcsec extended  galaxy\\
   & & & & & 0.02 & 0.01 & 0.01 & 0.01 & 0.05 & $\pm$0.035 & $\pm$0.071 & & &\\
WISE J205847.91--573653.8 & ORC~2~C & 0.2? & --- & --- & 21.38 & 20.95 & 20.82 & 20.72 & 20.37 & 15.499 & 14.930 & 11.729 & 0.569 & \alter{0.286$^1$} & edge-on spiral \\
& & & & & 0.02 & 0.02 & 0.02 & 0.03 & 0.2 & $\pm$0.041 & $\pm$0.066 & $\pm$0.232& & \\
\hline
\end{tabular}
\alter{Redshifts are taken from $^1$\citep{bilicki}, $^2$\citep{desphotoz}
\revb{$^3$\citep{zou19}}
}
\end{table*}

\begin{table*}
\centering
\tiny
\caption{Properties of the optical/IR source at the centre of ORC~4} 
\setlength{\tabcolsep}{3pt} 
\renewcommand{\arraystretch}{1.25} 
\label{tab:orc4-wise}
\begin{tabular}{cccccccccccccccl}
\hline
 & & GMRT & \multicolumn{2}{c}{GALEX} & \multicolumn{5}{c}{SDSS} & \multicolumn{3}{c}{WISE}\\
 Source Name & ID & flux & FUV & NUV & u & g & r & i & z & W1 & W2 & W3 & W1--W2 & z & Notes\\
 & & [mJy] & \multicolumn{2}{c}{[mag]} & \multicolumn{5}{c}{[mag]} & \multicolumn{3}{c}{[mag]}\\
\hline
 WISE J155524.65+272633.7 & G & 1.15 & ---  & ---  & 22.61 & 21.18 & 19.64 & 19.00 & 18.40 & 14.847 & 15.119  & 12.341  & --0.272 & 0.385 \\
SDSS J155524.63+272634.3 & & ? & & & $\pm$0.70 & $\pm$0.09 & $\pm$0.03 & $\pm$0.03 & $\pm$0.06 & $\pm$0.057 & $\pm$0.112  & $\pm$0.483 & $\pm$0.126 &  \\
\hline
\end{tabular}
\end{table*}

Galaxy classification was estimated using the WISE colours, following the WISE colour-colour diagram   \citep{wise},
and also using the DES and GALEX colours in a colour-magnitude diagram  \citep[e.g.][]{masters11, schawinski14}.
However these classifications are quite uncertain because \alter{they are redshift-dependent and} most of the galaxies do not have a spectroscopic redshift.

\section{Results}
\begin{figure*}
\begin{center}
\includegraphics[width=12cm]{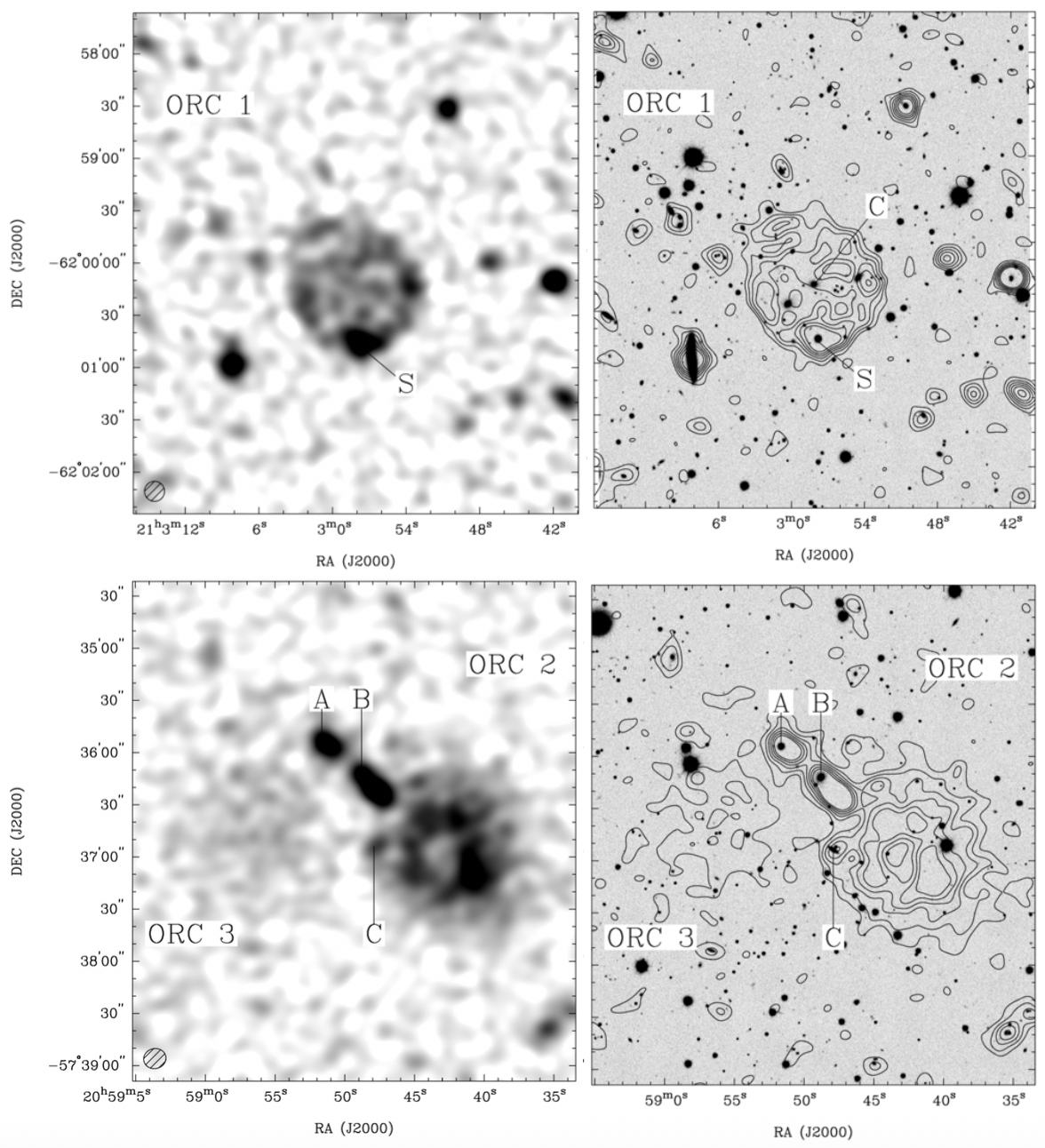}
\includegraphics[width=11.7cm]{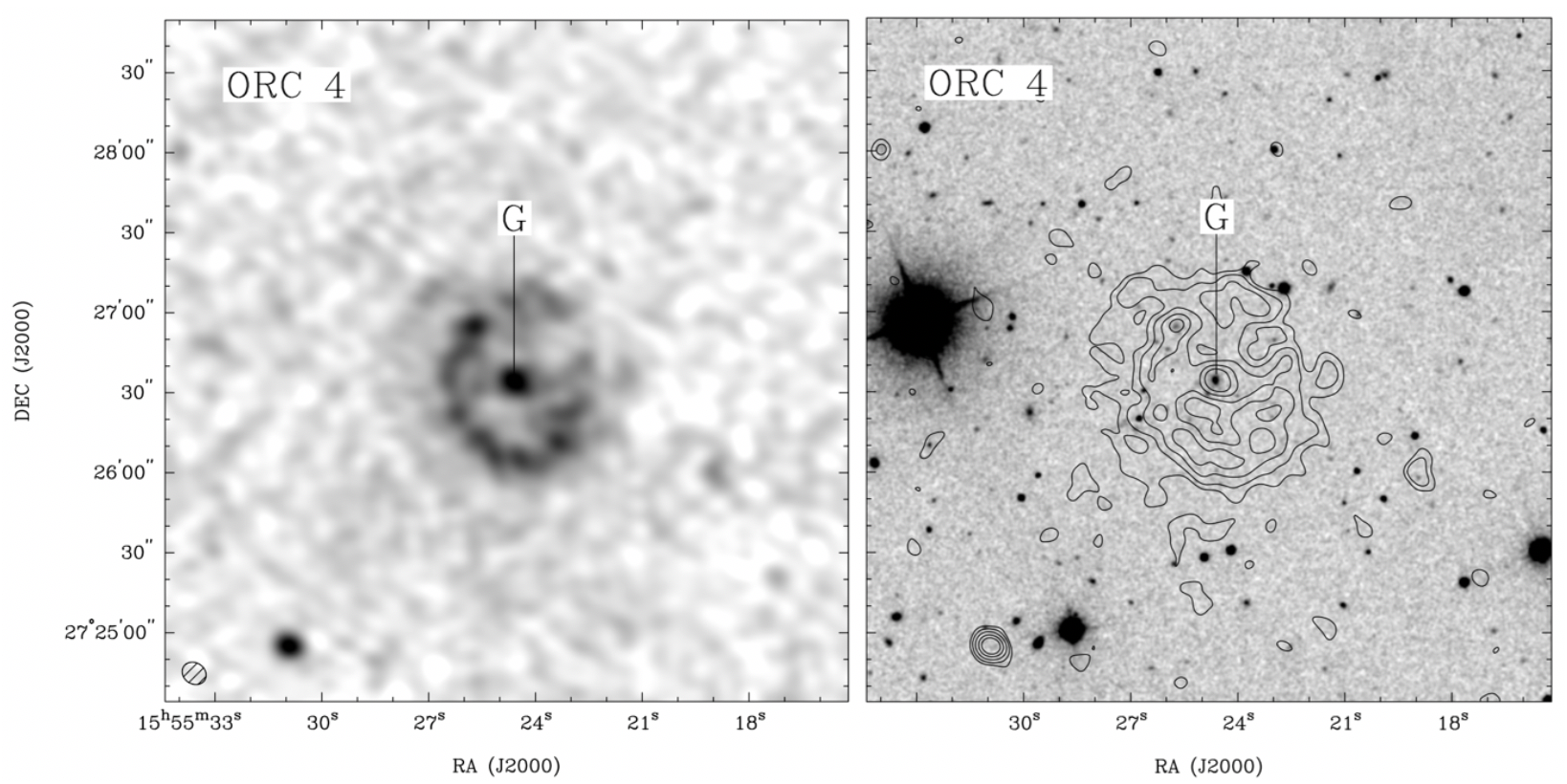}
\caption{ASKAP radio continuum images at 944 MHz of ORCs~1--3 from the EMU Pilot Survey  \citep{norris20}, and at 325 MHz of ORC~4 from GMRT archival data. On the left are greyscale images, with the  synthesized beam shown in the bottom left corner, and radio contours overlaid onto DES optical images on the right, as described in the text. The contour levels for ORC~1 and ORC~2 are 45, 90, 135, 180, 225, and 270 $\mu$Jy\,beam$^{-1}$, and contour levels for ORC~4 are 150, 250, 400, 600, and 800~$\mu$Jy\,beam$^{-1}$. Sources of interest are labelled (see Tables~3 \& 4). }
\label{fig:wtfimages}
\end{center}
\end{figure*}

Figure~\ref{fig:wtfimages} shows the radio and optical images of the ORC~1 (top), ORCs~2 \& 3 (middle) and ORC~4 (bottom), and their positions are listed in Table \ref{tab:list1}. Measured flux densities and spectral indices are shown in Table \ref{tab:orc1flux}.

\begin{figure*}
\begin{center}
\includegraphics[height=9cm]{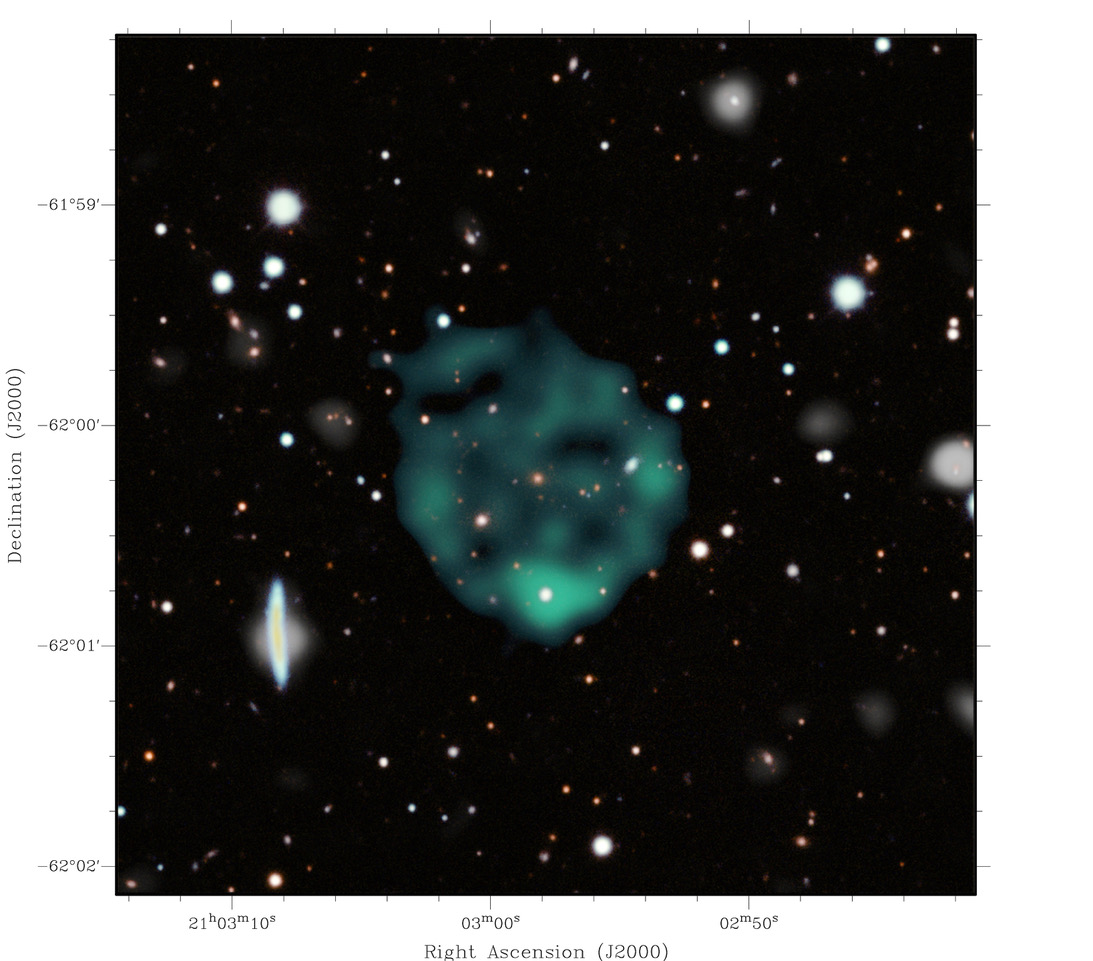}
\caption{An image of ORC~1 based on the same data as shown in Figure \ref{fig:wtfimages}, at native resolution, \rev{but enhanced to show faint features, particularly the internal structure or ``spokes'' of the ORC.} We made two images of the radio data using kvis: (a) an image using a square root transfer function to trace finer structure over the ORC region, and assigned the colour turquoise, and (b)  an image using a logarithmic transfer function  and assigned the colour blue-green. These two radio images were combined using the screen algorithm. For the optical data, the DES bands g, r, i, and z were assigned turquoise, magenta, yellow and red, respectively and combined using GIMP.  The optical/NIR image and the radio image were then combined using a masking technique. }
\label{fig:prettyorc1}
\end{center}
\end{figure*}

\begin{figure*}
\begin{center}
\includegraphics[height=9cm]{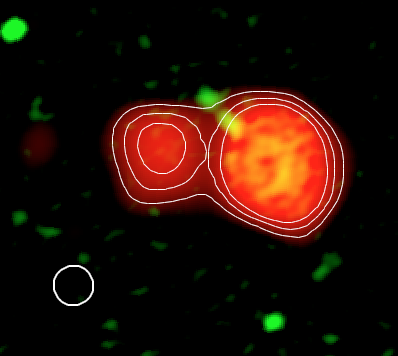}
\caption{An image of ORCs 2-3 based on the same data as shown in Figure \ref{fig:wtfimages}, \rev{but enhanced to emphasise the diffuse emission}. The green image is the original EMU-PS image at native resolution.  The red image and contours were produced by filtering out the small-scale emission using the multi-resolution technique described by \citet{rudnick02} with a filter size of 38 arcsec, and then convolving the residual with a Gaussian kernel of 40 $\times$ 40 arcsec full-width half-maximum (shown as the circle in the lower left of the image).}
\label{fig:prettyorc2}
\end{center}
\end{figure*}

It is difficult to appreciate the structure of these sources in a simple greyscale image with a linear transfer function, so in Figures \ref{fig:prettyorc1} and \ref{fig:prettyorc2} we show images that have been modified to bring out features of interest that may not be apparent in  Figure~\ref{fig:wtfimages}. The details of the transformation applied are shown in the Figure captions.

\begin{table*}
	\centering
	\caption{The new circular objects (``ORCs'') }
	
	\label{tab:list1}
	\begin{tabular}{cccccccl} 
		\hline
		ID & Name &  RA (deg) & Dec (deg) & l & b & \alter{ observation} \\
		   &      &  J2000    &  J2000  & (deg) & (deg) &   \\
		\hline
		ORC~1 & EMU~PD~J210357.9--620014 & 315.74292 & --62.00444 &333.41592   & -39.00906   &EMU-PS\\
		ORC~2 & EMU~PD~J205842.8--573658 & 314.67833 & --57.61611 & 339.08813  &  -39.52277 & EMU-PS\\ 
        ORC~3  & EMU~PD~J205856.0-573655&  314.73458 & --57.61528 & 339.08147   & -39.55247 & EMU-PS\\
      ORC~4 & 155524.63+272634.7 & 238.85272 &
+27.44271 & 44.35860 &   49.36566  & GMRT \\
		\hline
	\end{tabular}
\end{table*}

\begin{table*}
	\centering
	\caption{Integrated Flux Densities (in mJy) and spectral indices of the radio sources associated with  ORCs~1--3.  \alter{The quoted uncertainties include a contribution from the systematic uncertainty in the flux density scale. }
	Measurements at 88, 118, and 154 MHz are made with MWA. Measurements at 944 MHz are made with ASKAP.
		Measurements at 2121 MHz are made with ATCA. 
		MWA measurements of ORC~1 include the flux density of  source S, and MWA measurements of  
		ORC~2 include the flux density of  source C. However, these will have a negligible effect on the fitted spectral indices.
		}

		\label{tab:orc1flux}
	\begin{tabular}{lcccccc} 
		\hline
		source & 88  & 118  & 154 & 944 & 2121 & $\alpha$\\
		 & MHz & MHz & MHz & MHz & MHz &  \\
		\hline
 		ORC~1  & 105$\pm$16.5 & 69.5$\pm$8.6 & 38$\pm$6.0 & 6.26$\pm$1.25 & 2.29$\pm$0.23  & --1.17$\pm$0.04\\
		ORC~1(S) &   &.     &    & 0.60$\pm$0.12 & 0.15$\pm$0.03  & --1.71$\pm$0.35\\
		\hline
		ORC~2  &  28$\pm$14.4 & 25$\pm$6.8   & 14$\pm$5.3 & 6.97$\pm$1.39 & 2.31$\pm$0.23  & --0.80$\pm$0.08 \\
		ORC~2(A) &   &      &    & 0.46$\pm$0.10 & 0.46$\pm$0.05 & 0.0$\pm$0.34 \\
		ORC~2(B) &   &      &    & 0.76$\pm$0.15 & 0.66$\pm$0.07 & --0.17$\pm$0.22 \\
		ORC~2(C) &   &      &    & 0.19$\pm$0.05 & 0.07$\pm$0.03 & --1.23$\pm$0.36 \\
		\hline
        ORC~3    &   &    & $<$5 & 1.86$\pm$0.37  & $<$1.0     & --0.50$\pm$0.20 \\
		\hline
	\end{tabular}

\end{table*}

None of the ORCs has obvious optical, infrared, or X-ray counterparts to the diffuse emission, although in two cases there is an optical galaxy near the centre of the radio emission. 

\begin{figure*}
\begin{center}
\includegraphics[width=8cm]{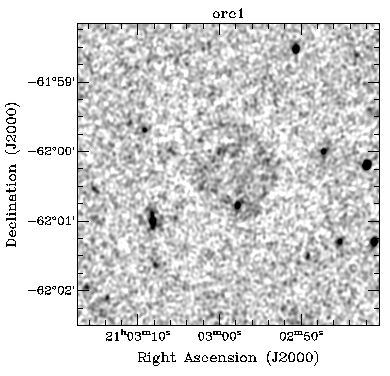}
\includegraphics[width=8cm]{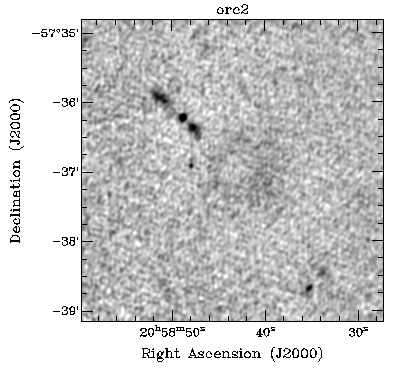}
\caption{ATCA radio continuum images of ORCs~1--3 at a frequency of 2.1 GHz. The image rms is about 12 \ujybm\ in both images. ORCs 1 and 2 are only faintly visible in these higher-frequency images, because of their steep spectral index and higher resolution, while ORC~3 is below the rms noise level. This image shows that sources A and B in ORC~2 are the two 
\alter{halves}
of an FRI radio galaxy.
}
\label{fig:atca}
\end{center}
\end{figure*}

ORCs 1--2 were subsequently observed at 2.1 GHz with the Australia Telescope Compact Array (ATCA) resulting in the images shown in Figure \ref{fig:atca}, and were also found in previously observed data at 88--154 MHz with the Murchison Widefield Array (MWA). 

Our current radio data do not enable the measurement of polarisation or in-band spectral index for such faint diffuse objects.

\section{Properties of the ORCs}

\subsection{ORC~1: EMU~PD~J210357.9--620014} 
\label{ORC1}

\begin{figure*}
\includegraphics[width=15cm]{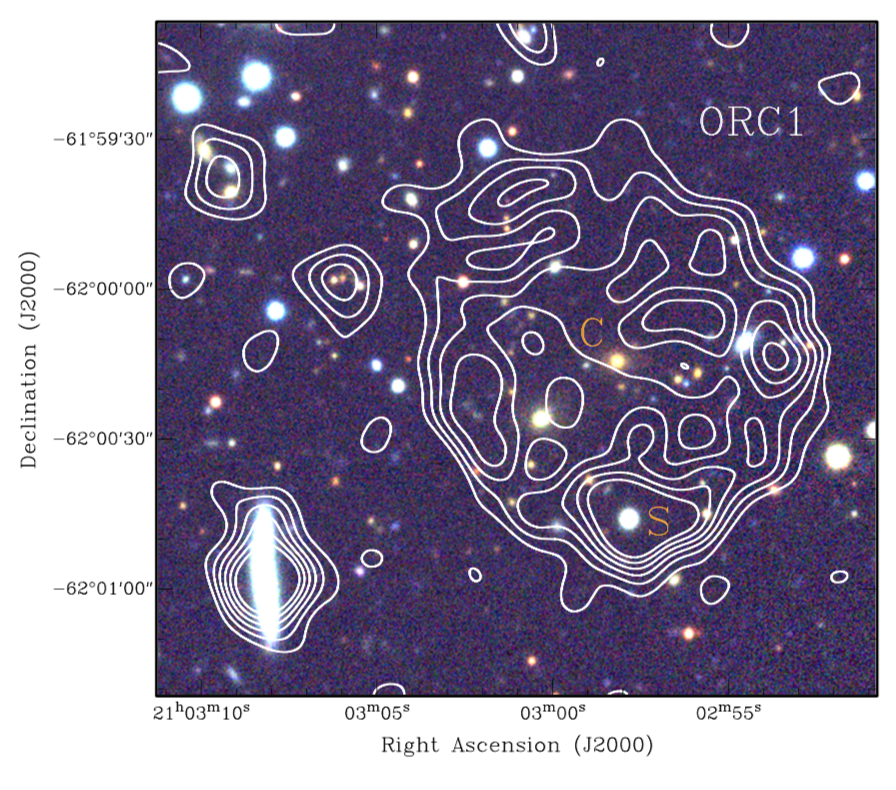}
\caption{ASKAP radio continuum image of ORC~1 (contours; see Fig.~1) overlaid onto a DES 3-color composite image; DES {\em gri}-bands are colored blue, green, and red, respectively. We identify two galaxies of interest: ``C" lies near the centre of ORC~1 and ``S" coincides with the southern radio peak (see Table~3).}
\label{fig:orc1-gri}
\end{figure*}

This 
radio  source (diameter $\sim$ 80 arcsec) has a near circular, edge-brightened filled morphology with brighter spots around its periphery. Figure~\ref{fig:orc1-gri} shows the radio contours overlaid on the DES 3-color ($gri$) optical image. 
There is no optical emission corresponding to the ring.
The typical radio brightness over ORC~1 in the ASKAP image is $\sim$ 130 \ujybm, with an integrated flux density of 6.26 mJy.  
By comparing our data from all the radio observations,  we derive a spectral index of the diffuse emission of $-1.17 \pm 0.04$. Such a steep spectral index may indicate
an ageing electron population, as often found in SNRs, cluster haloes, and  dying radio galaxies \citep{murgia}.

On the southern edge of the ring is a bright radio source 
(labelled ``S'' in \rev{Figure~\ref{fig:wtfimages}}) which is associated with a galaxy detected both by WISE (WISE J210257.88--620046.3)   \citep{wise} and by GALEX (GALEXASC J210257.91--620045.4)   \citep{galex}. The WISE colours indicate that this is a star-forming galaxy, and possibly a starburst or LIRG (luminous infrared galaxy). 
\revb{\citet{zou19} measure a photometric redshift of 0.228 for this galaxy.}

The radio observations listed in Table \ref{tab:orc1flux} show that source S has a spectral index of $\sim-1.71\pm0.35$, making it more likely to be an active galactic nucleus (AGN) than a star-forming galaxy.  The space density of radio sources in the EMU-PS  at least as bright as source S is $\sim390$ per \sqdeg, giving a $\sim$ 15\% probability that 
an unrelated source \alter{may be located within the $\sim$ 1.4 arcmin$^2$ area of ORC~1}.  
It is therefore possible that this is an unrelated object, physically unconnected to the diffuse object.

At the centre of the ring is a faint optical object (WISE J210258.15--620014.4), labelled ``C'' in \rev{Figure~\ref{fig:wtfimages}}, with no detectable radio emission in any of our observations. The enlarged DES $gri$-band image  in Figure~\ref{fig:orc1-gri} shows that this central object is extended E-W with a total extent of about 6 arcsec, (c.f. the $\sim$ 1~arcsec resolution of DES) and is probably a galaxy.

\begin{figure*}
\includegraphics[width=15cm]{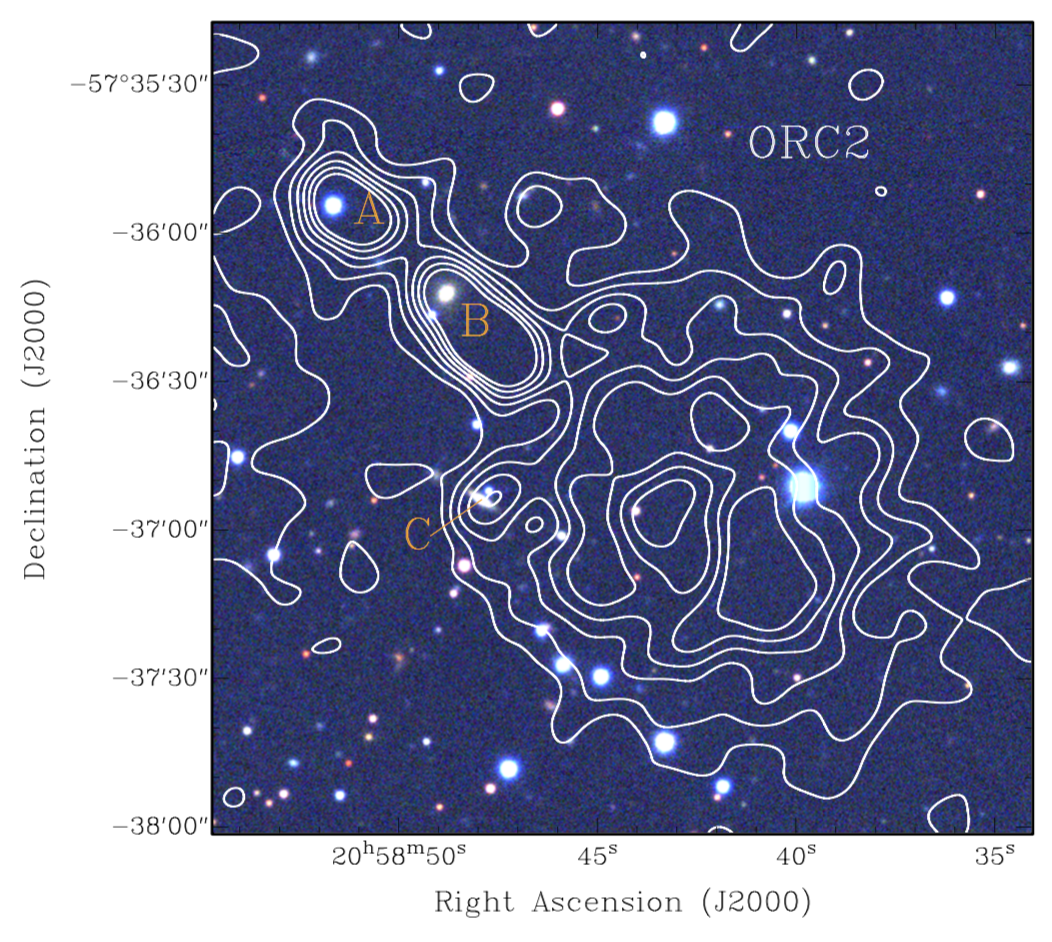}
\caption{ASKAP radio continuum image of ORC~2 (contours; see Fig.~1) overlaid onto a DES 3-color composite image; DES {\em gri}-bands are colored blue, green, and red, respectively. We identify three sources of interest, annotated A, B and C (see Table~4).}
\label{fig:orc2-gri}
\end{figure*}

\begin{figure*}
\includegraphics[width=15cm]{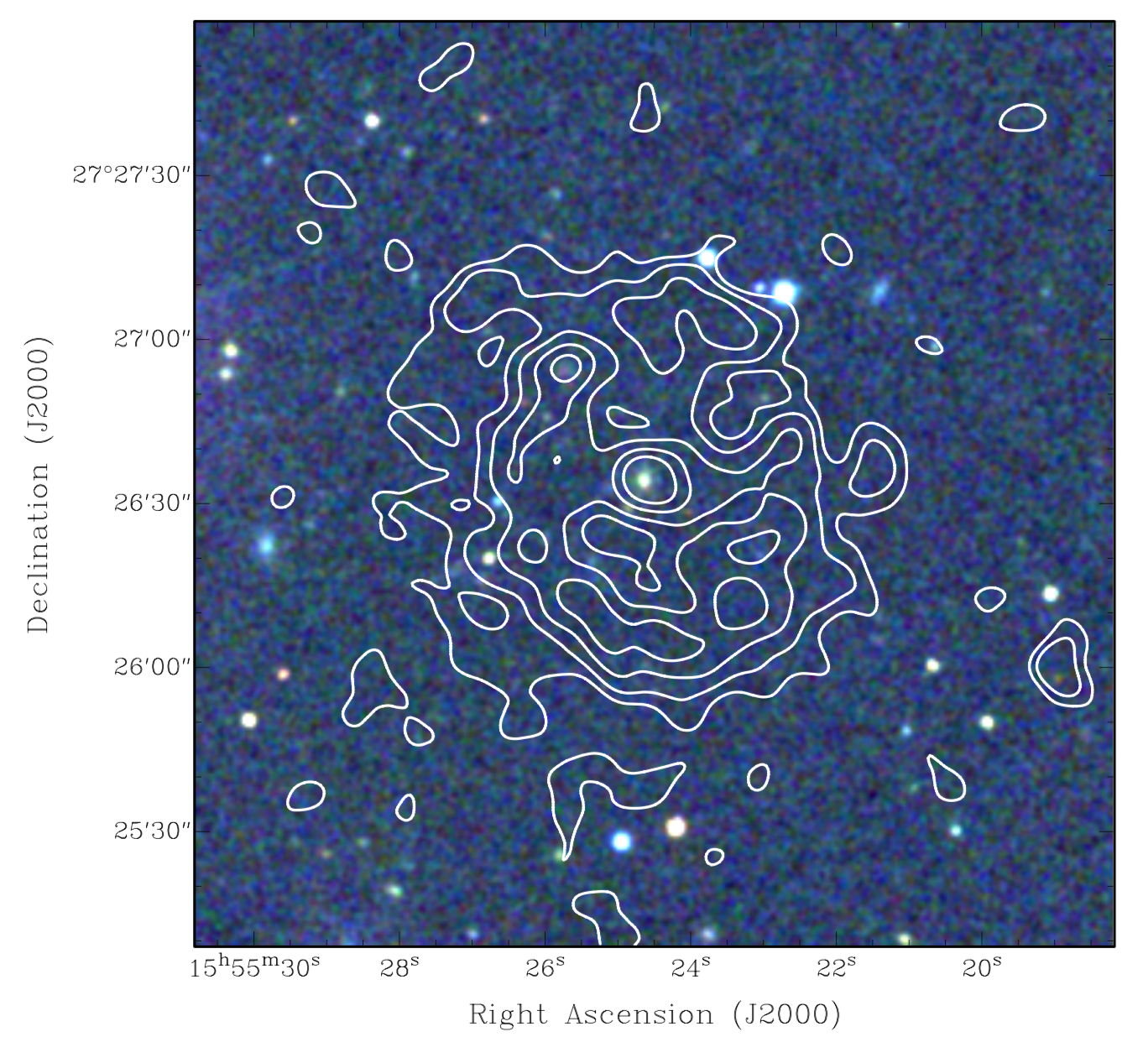}
\caption{GMRT radio continuum image of ORC~4 (contours; see Fig.~1) overlaid onto a SDSS 3-color composite image; SDSS {\em gri}-bands are colored blue, green, and red, respectively.}
\label{fig:orc4-gri}
\end{figure*}
The WISE colours for the central object (W1--W2 = 0.081, W2--W3 $<$ 2.045 ) do not enable us to distinguish between
a star-forming galaxy or a quiescent galaxy, but the optical colours suggest a quiescent galaxy.
\revb{\citet{zou19} measure a photometric redshift of 0.551 for this galaxy.}

\revb{At z=0.551, 6 arcsec corresponds to a diameter of 39 kpc,} making it unlikely to be a normal star-forming galaxy at high redshift.  The local space density of galaxies  (6333 per sq deg at W1$<$15) mean that there is a 15\%  probability of there being a source brighter than 15 mag within 10 arcsec of the centre.  
We therefore consider it possible that 
this is an unrelated object.

One arcmin to the south-east of the radio source  is a 16-magnitude galaxy (WISE J210308.23--620055.0) which is very elongated north-south in the optical image, and also appears as an elongated radio source. The WISE colours of this source are that of a spiral, so we consider it likely to be an edge-on spiral galaxy. It appears to be unconnected to the diffuse radio object and will not be discussed further.

\subsection{ORC~2: EMU~PD~J205842.8--573658.} 

Like ORC~1, this unusual radio source (diameter $\sim$ 80 arcsec) has a near circular, edge-brightened filled morphology with brighter spots around its periphery. Figure  \ref{fig:orc2-gri} shows the radio contours overlaid on the DES 3-color (gri) optical image.

The typical brightness over the diffuse object is $\sim$ 100--200 \ujybm, and the integrated flux density over the object (excluding sources A,B, and C) is $\sim$ 6.97 mJy at 944 MHz. 

To the north-east of ORC~2 in the ASKAP image is a pair of strong radio components. The ATCA image (Figure \ref{fig:atca}) shows a flat-spectrum source between them which is not visible on the ASKAP image, but which appears to be the central component of a double-lobed radio source, and which is within $\sim$ 1 arcsec of the  nearby optical galaxy labelled ``B'', which is  a 2.5 arcsec extended  galaxy, at a redshift of 0.311.

The space density of components in the EMU-PS at least as strong as source B is about 219 per \sqdeg, so the probability of finding B within 1 arcmin of the ORC is about 19\%, 
\rev{and so while it is tempting to consider that this source is associated with the ORCs, it is also possible that this source is an unrelated  object. We discuss a possible role for this source is Section \ref{RG}.
}

Within the eastern limb of ORC~2 in the ASKAP image is a compact radio source which is coincident with the galaxy labelled C in Table \ref{tab:orc1-wise}, which appears in the DES image to resemble an edge-on disk galaxy. \revb{Its  colour of g-r=0.43 confirms that it is a star-forming galaxy, and so we classify this galaxy as an edge-on spiral. \citep[e.g.][]{masters11}.} The space density of components in the EMU-PS at least as strong as source C is about 1000 per \sqdeg, so the probability of finding B within 1 arcmin of the ORC is about 90\%, and so we consider it likely that this source is an unrelated  object.

\subsection{ORC~3: EMU~PD~J205856.0-573655.} 

Immediately to the east of ORC~2 is a another faint circular patch, which is visible in two independent observations with ASKAP,  but is too faint to be seen in ATCA or MWA data.  This faint diffuse patch, which we call ORC-3, has a typical brightness of $\sim$ 50--80 \ujybm, and the integrated flux density over the object is $\sim$ 1.86 mJy at 944 MHz. It has a spectral index of $\sim -0.50$, although this is quite uncertain because it depends on
\alter{one detection bracketed by two upper limits}.
 
ORC~3 appears to be a fainter example of the ORC phenomenon exhibited by ORCs~1 and 2, but it is puzzling that it is so close to ORC~2. Within the EMU-PS, there are 2 very obvious ORCs (ORCs 1 and 2) and about six fainter ``candidate ORCs'', \alter{found during a careful visual inspection of the area of the EMU-PS. While this is not a rigorously defined statistical sample, it enables us to estimate} a space density of \alter{$\lesssim$} 0.03 per \sqdeg. The probability of one of these lying within 2 arcmin is therefore \alter{$\lesssim$} $10^{-4}$. We therefore consider it unlikely that this is a chance association, but instead deduce that ORC~2 and ORC~3 are in some way related.

\alter{Assuming that ORC2 and 3 are driven by the same mechanism as the other ORCs, this militates against the ORCs being caused by a transient event.}

\subsection{ORC~4}

\rev{
ORC~4 was discovered serendipitously by Intema in archival observations  \citep{venturi17} of the cluster Abell 2142 taken at 325 MHz with the Giant MetreWave Radio Telescope   \citep[GMRT:][]{gmrt} in March 2013. \alter{Note that Abell 2142 is 14 arcmin away from ORC4, at a redshift of 0.091, so is unlikely to be physically associated.} Intema viewed hundreds of GMRT images before finding ORC4, suggesting these sources are rare. 
}
In most respects ORC~4 is very similar to ORCs 1--3, but differs in having a central radio continuum source.

ORC~4 is marginally detected at 150 MHz in TGSS   \citep{intema17} and at 1.4 GHz in NVSS   \citep{nvss}, but in both cases the  image quality is poor because of  poor sensitivity, uv-coverage, and low resolution. The measured total flux densities are given in Table \ref{tab:orc4flux}.

\begin{table}
	\centering
	\caption{Measured Flux Densities (in mJy) of the radio sources associated with  ORC~4. 150 MHz data are from TGSS  \citep{intema17}, 325 MHz data are from the observations described in the text, and 1400 MHz data are from the NVSS survey  \citep{condon98}. The flux densities for the ring are the total flux densities (including source G) measured in a 2 arcmin diameter aperture, and that for source G is from a fitted Gaussian component.
	}
	\label{tab:orc4flux}
	\begin{tabular}{lcccc} 
		\hline
		source    & 150     & 325          & 1400      & $\alpha$\\
		          & MHz     & MHz          & MHz       &    \\
		\hline
		ORC~4     & 39$\pm$10 & 28$\pm$2.8     & 5.3$\pm$0.7 & --0.92$\pm$0.18
		\\
		ORC~4(G)  &         & 1.43$\pm$0.13  &           &  \\
		\hline
	\end{tabular}
\end{table}

Combining all aperture flux measurements, we fit a single spectral index 
$\alpha$ = --0.92 $\pm$ 0.18.

ORC~4 differs from the other ORCs in having a central radio source, labelled ``G'' in Figure \ref{fig:wtfimages}, which is listed in the NVSS catalogue   \citep{nvss} as NVSS J155524+272629. 
It is coincident with a red galaxy seen in both SDSS and Pan-Starrs images, and detected by WISE. The compact source has an unambiguous optical/IR counterpart in SDSS (SDSS J155524.63+272634.3) and WISE (WISEA J155524.65+272633.7) with a photometric redshift of 0.39   \citep{bilicki} implying a linear size of the ring of \rev{410 kpc $\times$ 350 kpc}.

\subsection{Summary of ORC properties}
\begin{figure*}

\includegraphics[width=7cm]{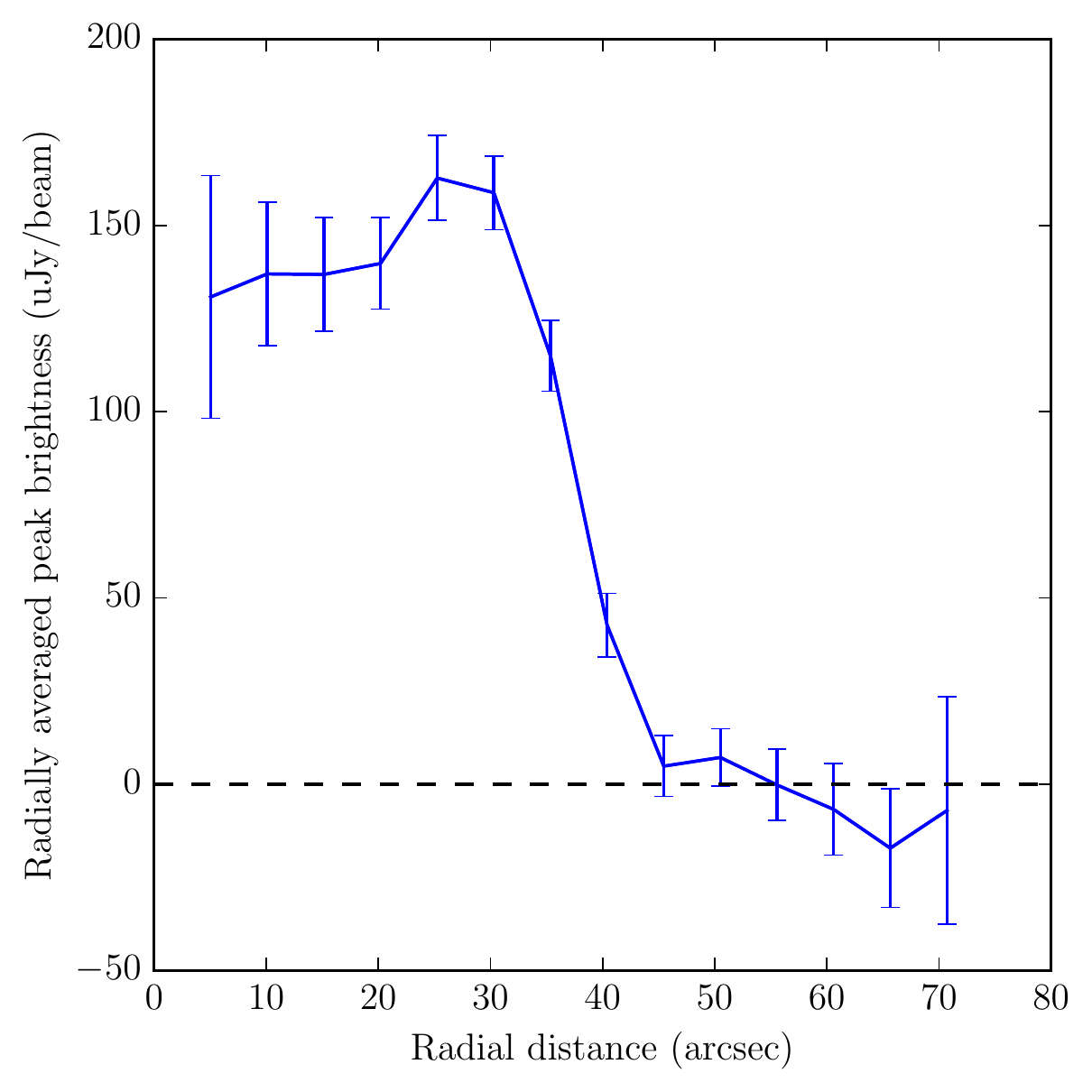}ORC~1
\includegraphics[width=7cm]{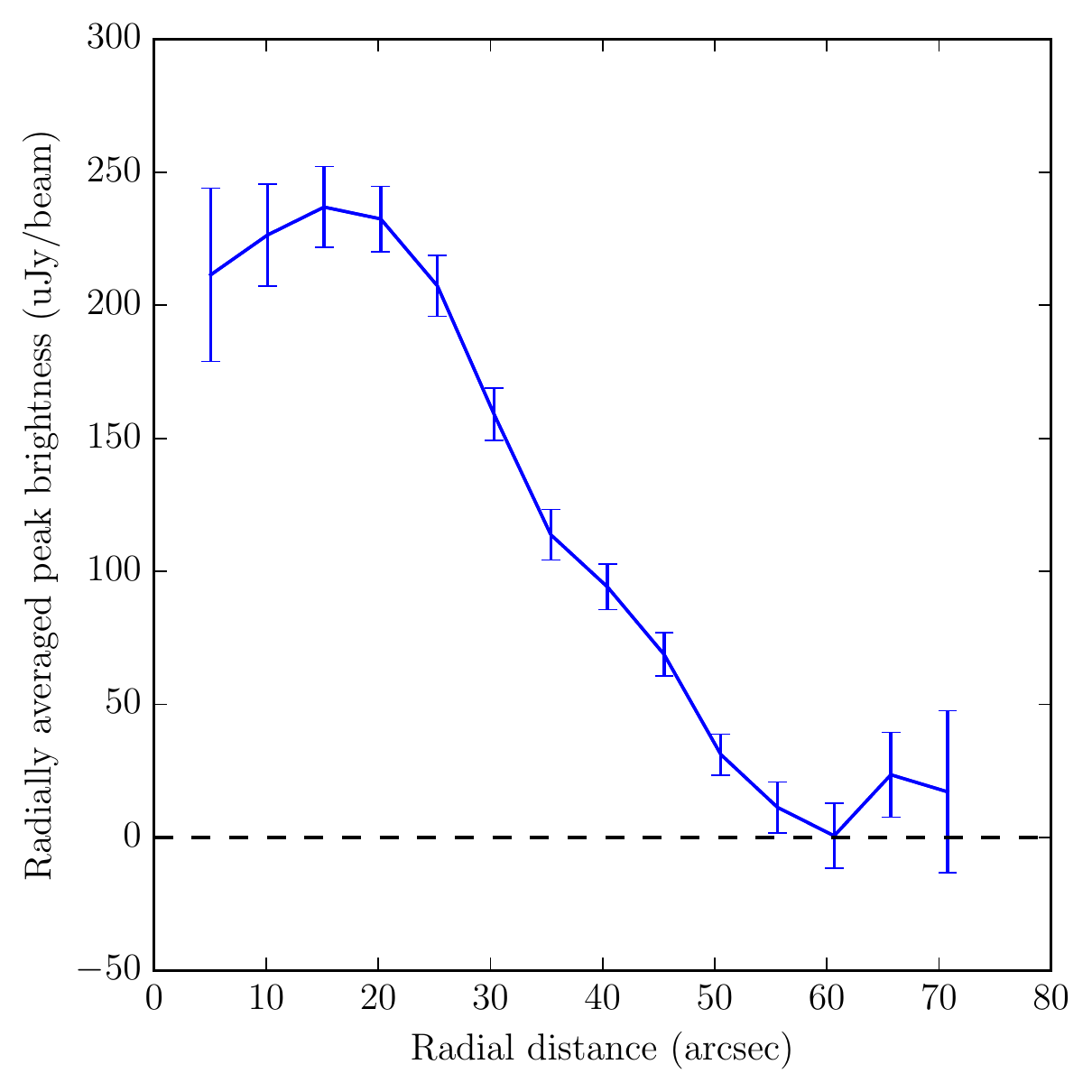}ORC~2
\includegraphics[width=7cm]{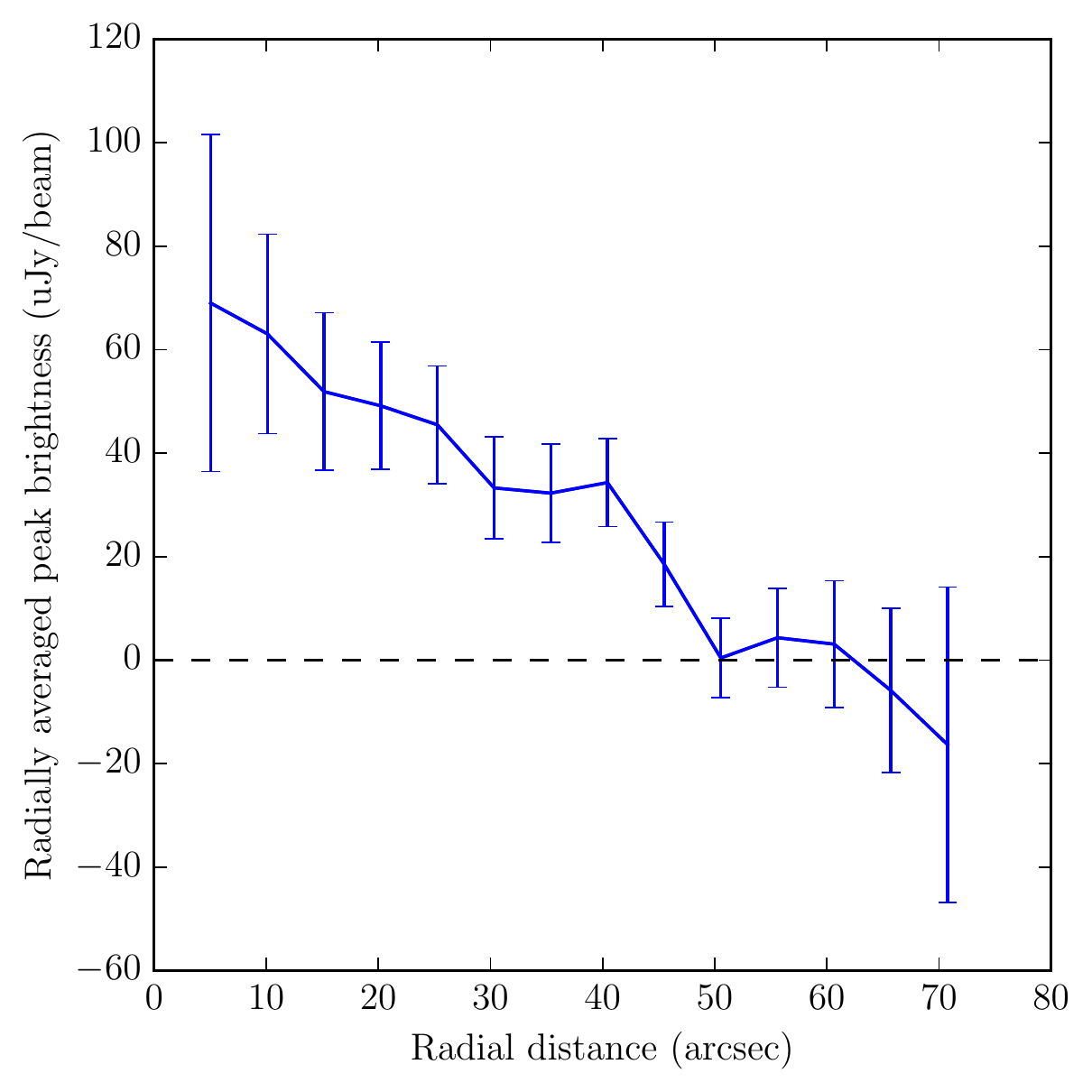}ORC~3
\includegraphics[width=7cm]{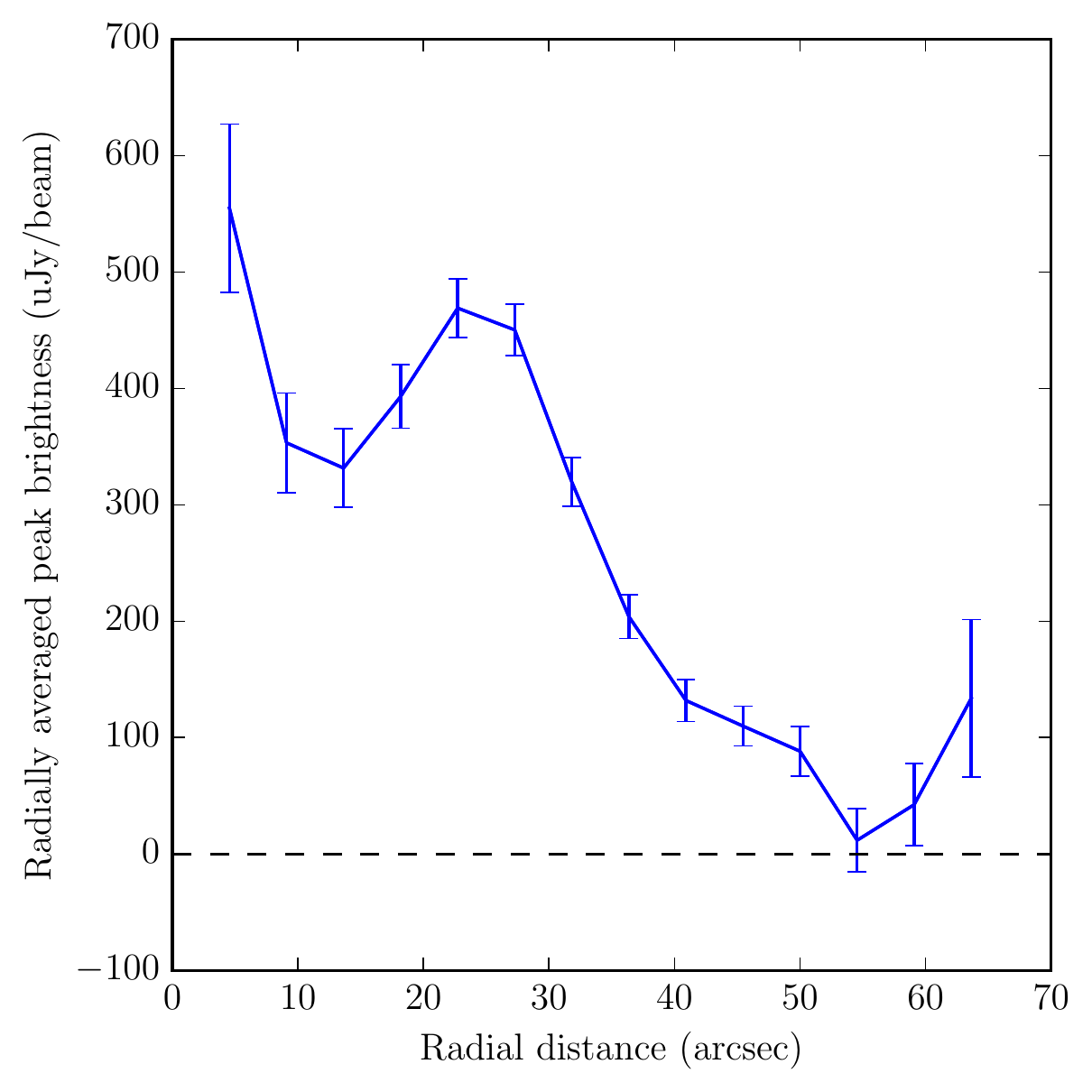}ORC~4
\caption{Radial profiles of the diffuse emission of the ORCs, measured from the ASKAP and GMRT data, and integrated radially around the ORC, assuming circular symmetry, after removing compact sources  A and B in ORC~2.  Error bars are $\sigma$ / sqrt (number of independent beam volumes), where $\sigma=25$\,\ujybm for the EMU data (ORCs~1, 2, and3), and 60\,\ujybm for the uGMRT data (ORC~4).}
\label{fig:profiles}
\end{figure*}

The four ORCs discussed here are similar in displaying a strong circular symmetry. They are also similar in  (a) having a diameter about 1 arcmin, (b) having a steep spectral index $\alpha \sim -1\,$ (c) being at high Galactic latitude. They differ in that (a) two of them have a central galaxy while two do not, and (b) three of them (ORCs 1, 2 \& 4) consist of a partly filled ring while one (ORC~3) seems to be a uniform disc. The radial profiles of the ORCs are shown in Figure \ref{fig:profiles}. ORC~1, ORC~2, and ORC~4 are similar in having a filled but edge-brightened disc, while ORC~3 decreases monotonically from the centre.

There is also the puzzling fact that two of them (ORCs 2 and 3) are very close together, implying that these two ORCs have a common origin. \rev{No such counterpart is visible for ORCs 1 and 4 above the noise in the image, implying that any such counterpart must be weaker than $\sim$10\% of the visible ORC. A counterpart similar to ORC 3 would be visible in both cases. We cannot of course exclude a counterpart hidden behind the visible ORC, and note the faint extended emission slightly to the east of ORC4 in Figure \ref{fig:wtfimages}, which might be interpreted as a second ring behind the first.}

If the central galaxy in ORC~4 is associated with the ring, then the ring is at a redshift of 0.39 and has a size of \rev{410~$\times$~350~kpc}.

We estimated proper motion by comparing the positions of ORCs 1 and 2 in the ASKAP (taken in November 2019) and ATCA observations (taken in March 2020). This is difficult because of their diffuse nature and the low SNR of the ATCA observations, but we estimate an upper limit of about 4 arcsec on any spatial shift of the diffuse emission between the two sets of observations, which rules out a solar system object.

\section{Discussion}
\alter{Our argument that these are new types of object naturally provokes two questions: (a) Are they real? and (b) are they new? 

We propose that at least ORCs 1 and 2 are certainly real because they have been independently detected by two different telescopes using different software, which produce images of rings with the same diameter. 

We propose that they are new because nowhere in the literature can we find an example of an object with the striking circularity of the ORCs, except for starburst rings and the Galactic objects which we argue below are not the cause of the ORCs. }

We now consider \rev{whether these objects might be explained by  known phenomena}. For the purposes of this discussion, we assume all four to have a similar cause, although we acknowledge that it is possible that we may have more than one type of object represented in the class of ORCs. \alter{For example, it could be that the single ORCs are produced by a transient event in a galaxy, while the double ORC is produced by a binary phenomenon such as the two jets from an AGN, even though no previously known double-radio-lobed AGN resembles ORC~2 and 3.}

\subsection{Imaging Artefact}
Circular artefacts are well-known in radio images, and are often caused by imperfect deconvolution of a strong source, resulting in some fraction of the telescope point-spread-function appearing in the final image. However, ORCs~1 and 2 are clearly imaged with two different telescopes, at different times, with different processing software, and all ORCs have been detected by more than one telescope. We therefore consider artefacts to be a very improbable explanation.

\subsection{Supernova Remnant}
\label{snrsection}

The morphology of the ORCs is remarkably similar to some typical supernova remnants (SNRs)   \citep{anderson17,2017ApJS..230....2B,green19,joseph19,2019A&A...631A.127M}. 
We calculate the probability that the ORCs are SNRs as follows.

\rev{
First we note that the 294 known SNRs\footnote{Green D. A., 2019, `A Catalogue of Galactic Supernova Remnants (2019 June version)', available at \url{http://www.mrao.cam.ac.uk/surveys/snrs/}}\citep{green19}
have a Galactic latitude distribution, shown in Figure \ref{fig:SNR}, which is strongly peaked at $ b \sim 0$, but with outliers extending out to b = --21.5\degr\ and b = +15\degr. The Galactic latitude of the four ORCs discussed here are --39.0\degr, --39.5\degr, --39.6\degr, and +49.4\degr, all of which are well outside the range of known SNRs, and so it seems unlikely that these four objects are members of this population.

\begin{figure}
\includegraphics[width=8cm]{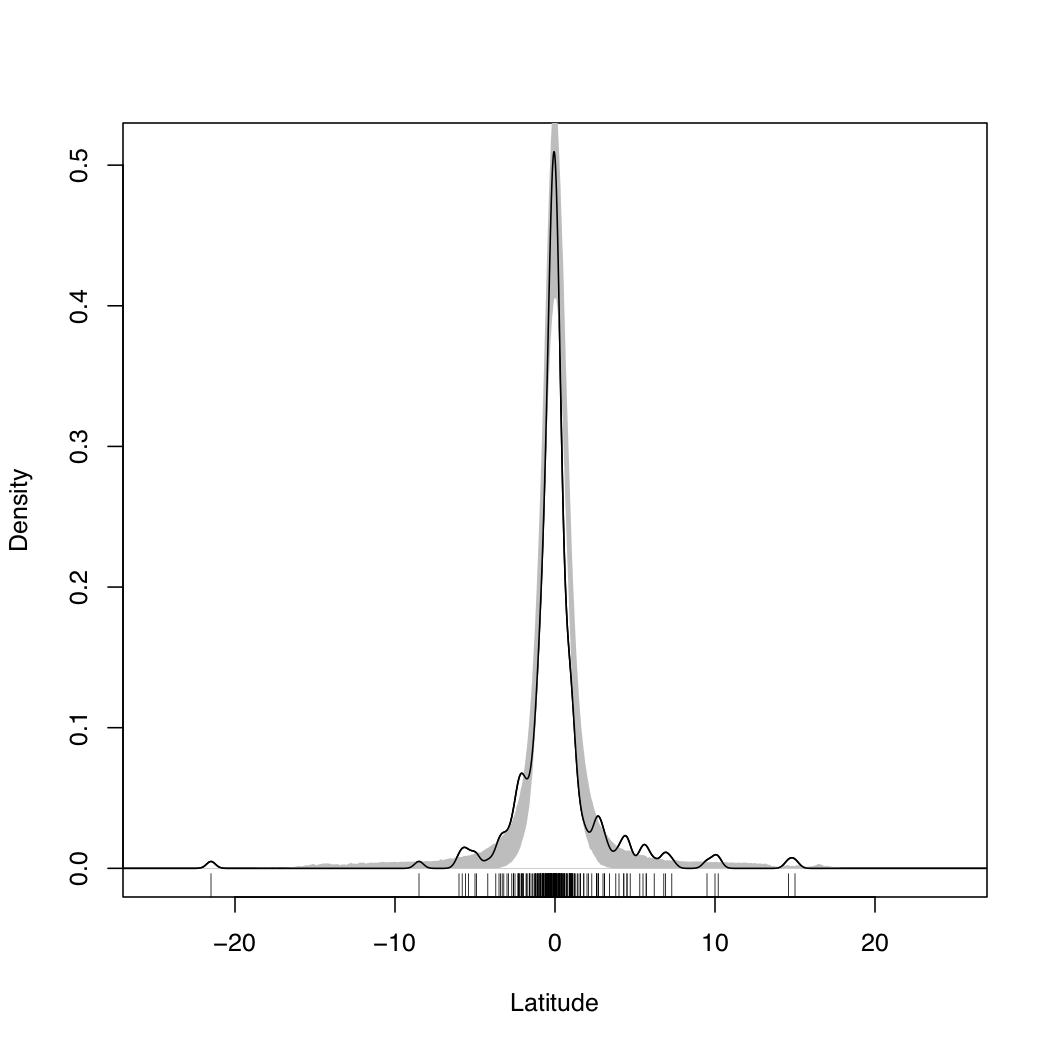}
\caption{\rev{The probability density of the Galactic latitudes of known SNRs. The lines at the bottom show the measured latitudes, taken from \citep{green19}, and the solid curve above shows them convolved with a Gaussian kernel density estimator, to show the distribution more clearly. A SNR at a Galactic latitude of --39.5\degr\ would be well outside the range of all known SNRs, and outside the range of this plot.
The grey band shows the  95\% confidence interval of  probability density  predicted by our model, which can be seen to be a good fit to the data. 
}}
\label{fig:SNR}
\end{figure}

To quantify this, we use a subset of SNRs with measured distances \citep{vukotic19} to construct a model of the SNR distribution in the Galaxy. The distribution of z values  (the distance from the mid-plane) of these SNRs is roughly Gaussian with a mean of 3.7 pc and a standard deviation of 127~pc. We then simulate a population with this distribution of z values, and a uniform distribution of x and y coordinates, within the cylinder of the Galaxy bounded  by the  solar circle which has a radius of 8.2~kpc \citep{gravity19}. We then measure the Galactic latitude of this population as observed from Earth, and show the resulting distribution overlaid on the real data in Figure~\ref{fig:SNR}. From these simulations we can calculate the probability of obtaining SNRs within the solar circle at latitudes -39.0\degr, -39.5\degr, -39.6\degr. 

 We find that the probability that at least one of these is a member of the parent population of SNRs is 0.055, and that the probability of all three belonging to the parent population of SNRs is  $2.1 \times 10^{-5}$. We therefore consider it unlikely that these are SNRs, although we cannot rule out the possibility that we have discovered a new population of high-latitude SNRs.
}

\subsection{Galactic Planetary Nebula}
Planetary nebulae (PNe) can also appear as diffuse disks of radio emission   \citep{meixner96}. 

\revb{3709 candidate PNe are currently listed in the HASH catalog \citep{parker17}, of which 2627 are ``true'' PNe (i.e. PNstat=T), of which 2226 lie within the solar circle.  Unfortunately very few reliable distances are known, precluding the construction of a model as we used for SNRs. Instead we model the angular distribution of the 2226 PNe. At the Galactic latitude of the ORCs (-30\degr < b <-50\degr), the HASH catalog  lists 3 PNe, giving a sky density of one PNe per 919 square degrees, so that the expected number of PNe in the 270 sq. deg EMU-PS is $\sim$ 0.3, giving a probability of $\sim$0.3 of finding one in the EMU-PS field, or $\sim$0.1 of finding three in the EMU-PS field. Thus PNe seem unlikely as a possible mechanism for ORCs, but cannot be completely excluded on the basis of the latitude distribution alone.

The radio emission of PNe is generated by thermal free-free emission, and is therefore expected to have a positive spectral index $\alpha$ \citep {hajduk18}, which might vary from $\alpha \sim$--0.1, in the optically thin case to $\alpha \sim$ +2 in the optically thick case. \citet{gruenwald07} and \citet{hajduk18} show a good fit between these calculated spectral indices and observations. This is quite different from the steep spectral indices for the ORCs listed in Tables \ref{tab:orc1flux} and \ref{tab:orc4flux}.

A wider range of spectral indices  are  reported in two surveys   \citep{2011MNRAS.412..223B, wang18}. However, both these surveys  suffer from large uncertainties because of poor frequency coverage and mismatched instrumentation. 

}

We therefore consider it unlikely that the ORCs are PNe.

\subsection{Ring around Wolf-Rayet star}
Wolf-Rayet (WR) stars can eject bubbles of material that appear at both radio and optical wavelengths as a ring of emission \citep{WR2001}.  \rev{667 WR  stars are known in our Galaxy\footnote{\url{http://pacrowther.staff.shef.ac.uk/WRcat}} \citep{rosslowe15}}, although it is estimated that there could be as many as 2000. However, the radio emission associated with WR stars is typically of size a few arcsec or less  \citep{abbott86} and they generally have a flatter spectral index than the ORCs  \citep{dougherty00}. \alter{The 667 known WR stars have Galactic latitudes ranging from -10.1\degr to +6.5\degr\ so the probability of finding one at a latitude of $\sim$ -39.5 is very small.} We therefore consider WR stars to be an unlikely cause of the ORCs.

\subsection{Face-on star-forming galaxy or ring galaxy}
\begin{figure}
\begin{center}
\includegraphics[width=8cm, angle=0]{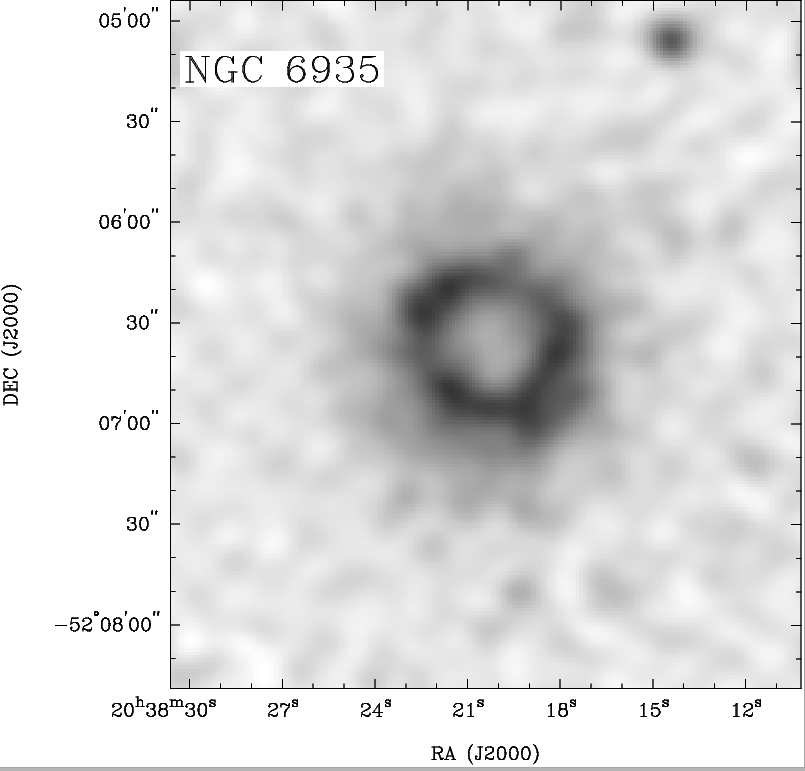}
\caption{ASKAP 944 MHz radio continuum image of the face-on, star-forming galaxy NGC~6935 ($v$ = 4543\kms), as observed in the EMU-PS. \alter{The ring is about 2 kpc in diameter. } }
\label{fig-sf}
\end{center}
\end{figure}

Ring-shaped star-forming galaxies such as the Cartwheel galaxy are well-known   \citep{higdon96}, and some nearby, nearly face-on spiral galaxies have star-forming rings, typically detected in the H$\alpha$ emission line and in radio continuum   \citep{pogge93,forbes94}. An example detected by ASKAP is shown in Figure \ref{fig-sf}. However, in all known cases these ring galaxies and starburst rings are bright at optical wavelengths, which contrasts with the lack of measurable optical emission from the ORCs on a similar scale to the radio emission. If the ORC emission corresponded to the size of a typical disk galaxy ($\sim$10--20\,kpc) it would lie at a distance of about 25--50\,Mpc ($z\sim 0.01$). The emission from such a large, nearby face-on disk galaxy would be easily detectable in the DES imaging. We therefore do not consider these to be a likely explanation for ORCs.

\subsection{Galactic Wind Termination Shock}
The winds from star-forming galaxies create a bubble surrounded by a termination shock. For a Milky Way-like galaxy forming $\sim$few \Msun/year in an isotropic environment, a roughly spherical galactic wind termination shock at a distance of $\sim$ (few - 10) $\times$ 10 kpc is predicted 
  \citep{jokipii87,volk04}; 
to reproduce the observed $\sim$1 arcmin angular diameter 
scale of an ORC, a galaxy 
with a termination shock at 30 kpc 
would then need to be located at an angular diameter distance of $\sim 100$ Mpc or $z  \sim 0.02$.

This shock at velocity $v$ will accelerate cosmic ray electrons to an energy limited by inverse Compton cooling of

$$
E_{\rm e,max} \sim 10^{13} \ {\rm eV} \ \left(\frac{B}{0.1 \ \mu {\rm G}} \right)^{-1/2}
\left(\frac{v}{500 \ {\rm km/s}} \right)
$$
where we have normalised $v$ to a conservatively small characteristic flow speed and a 0.1 $\mu$G field   \citep{crocker}.
This implies that the termination shock associated with a star-forming galaxy  
should be easily capable of accelerating CR electrons to the $\sim$ few $\times$ 10 GeV energies at which they would produce synchrotron radiation at $\sim$1 GHz.

The energetics of this scenario are also reasonable: assuming that non-thermal electrons account for 1\% of the mechanical power dissipated at the putative shock, a mJy source located at $z \sim 0.02$ requires a shock dissipating $\sim 10^{36-37}$ erg/s which can be easily energised by a host with a star formation rate of a few solar masses per year.

While this is a theoretical possibility, such a shock has not yet been observed elsewhere.

\subsection{A bent-tail radio galaxy}
\begin{figure}
  \begin{center}
  \includegraphics[width=8cm, angle=0]{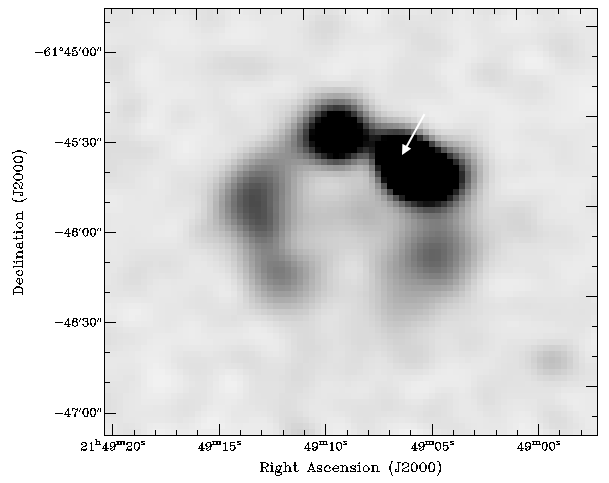}
  \caption{EMU-PS image of the bent-tail radio galaxy EMU~PD~J214905.4-614542. The position of the host galaxy is indicated by an arrow. \alter{No redshift is available for this source.}}
  \label{fig:circlewat}
  \end{center}
\end{figure}

In a bent-tail radio galaxy, the two jets are bent/curved by their relative motion through the intra-cluster medium.  In the case of the source shown in Figure~\ref{fig:circlewat}, the bending is so severe as to form almost a circle, and its formation would require additional forces or jet variations over and above relative motion.  In the case of ORC~1, it could be argued that source S is the host galaxy.  However, jets from SF galaxies are rare, and bent ones have not yet been seen. Similar to the ORCs, there is faint central emission between the bent tails, approximately 30\% as bright as the tails. However, no bent-tail galaxy, including the source shown in Figure~\ref{fig:circlewat}, shows the striking circular symmetry of the ORCs, and so we do not consider this to be a likely cause of the ORCs.

\subsection{Lobe from a double-lobed radio galaxy, viewed side-on}
\label{RG}
\begin{figure}
\begin{center}
\includegraphics[width=8cm, angle=0]{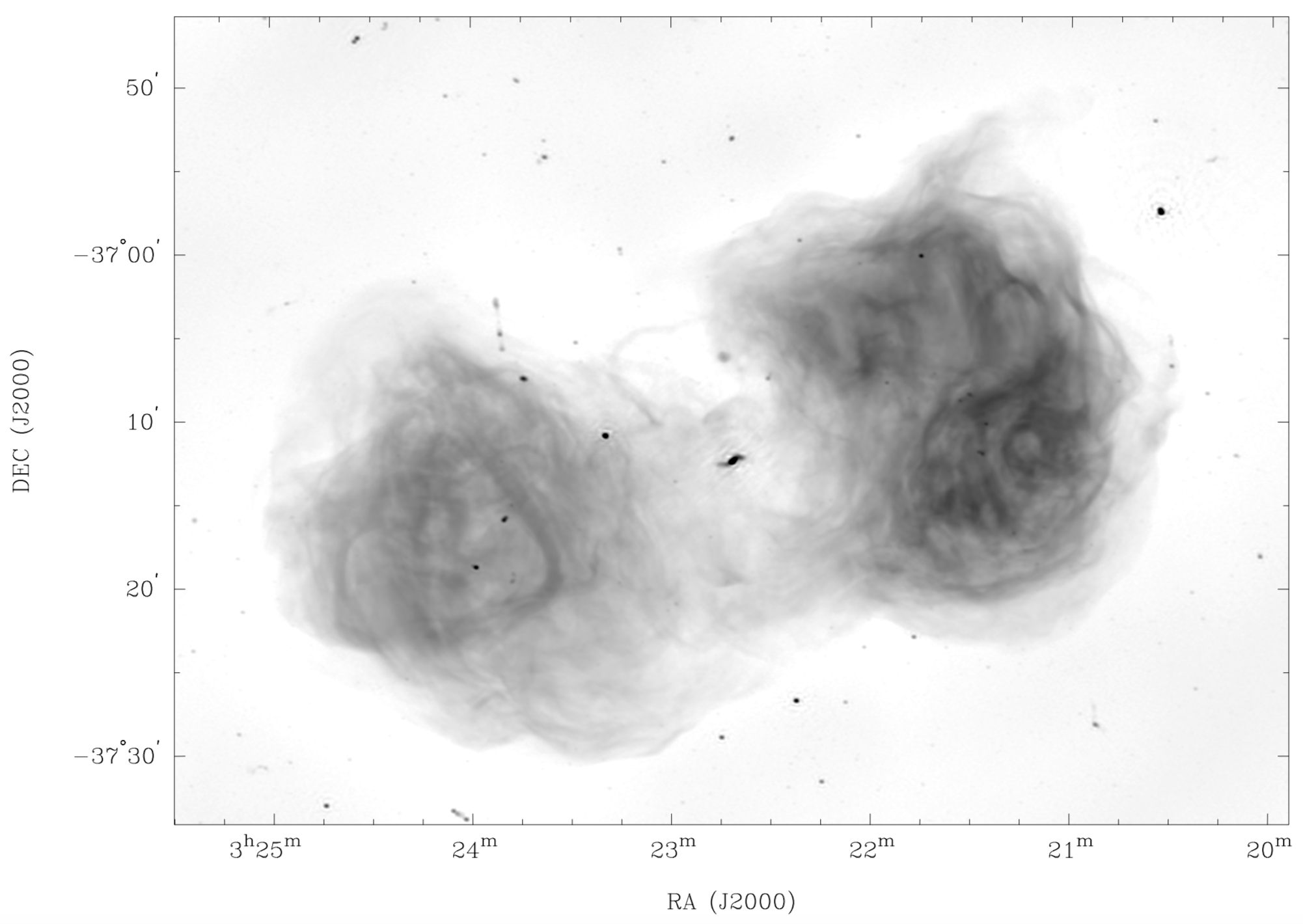}
\caption{ASKAP 944 MHz radio continuum image of the double-lobe radio galaxy Fornax A, from unpublished ASKAP data.\alter{The largest angular size across the lobes is about 30 kpc.}}
\label{fig:fornaxa}
\end{center}
\end{figure}

The ORCs might, in principle, be one lobe of a double-lobed radio galaxy. For example, the radio galaxy Fornax A has two near-circular radio lobes, shown in Figure~\ref{fig:fornaxa}. However, we consider this unlikely \alter{ primarily because the ORCs are strikingly circular, and in most cases edge-brightened, unlike the morphology typically seen in double-lobed radio galaxies.}

\rev{
ORCs 1 and 4 also have no corresponding lobe, although it is possible that a corresponding lobe may be below our sensitivity threshold.

Because ORCs 2 and 3 are close together,  it is tempting to consider them as the lobes of a radio galaxy. However, there is no sign of a central optical or radio host source between the two lobes. Source C in the limb of ORC~2 could potentially be a host, but it is an edge-on star-forming galaxy, which rarely host double-lobed radio sources.

A possible host is the  double-lobed AGN (sources A and B in Figure 1) close to ORCs 2 and 3.  However, although the AGN points roughly at ORC2, it does not point at ORC3, and it is not located between them, but to the north. 
}

\subsection{Lobe from a double-lobed radio galaxy, viewed end-on}
\label{RG2}
If a radio-loud AGN is viewed ``down the barrel'' of the jet,  the  end-on radio lobe can appear as a \alter{roughly} circular object, as seen in some BL Lac sources   \citep{ulvestad}. If the central radio source were precessing, then the central spot could in principle be a circle, although this has not yet been observed. However, such sources are accompanied by (in the case of a BL Lac source) a bright, blue, unresolved (sub-arcsec) optical counterpart, or (in the case of a radio galaxy) by a quiescent galaxy. \revb{In such sources, the central radio emission is brighter than the fainter halo, and this is not seen in the ORCs}.

\alter{This model is also inconsistent with}
the inferred physical sizes. If the ORCs are at redshifts at $z \sim$ 0.3, which is consistent with faint optical counterpart galaxies and with the measured redshift of source G in ORC~4, then the transverse sizes would be of order of 400 kpc, which is about an order of magnitude greater than observed in other radio galaxies.

\begin{figure}
\begin{center}
\includegraphics[width=8cm, angle=0]{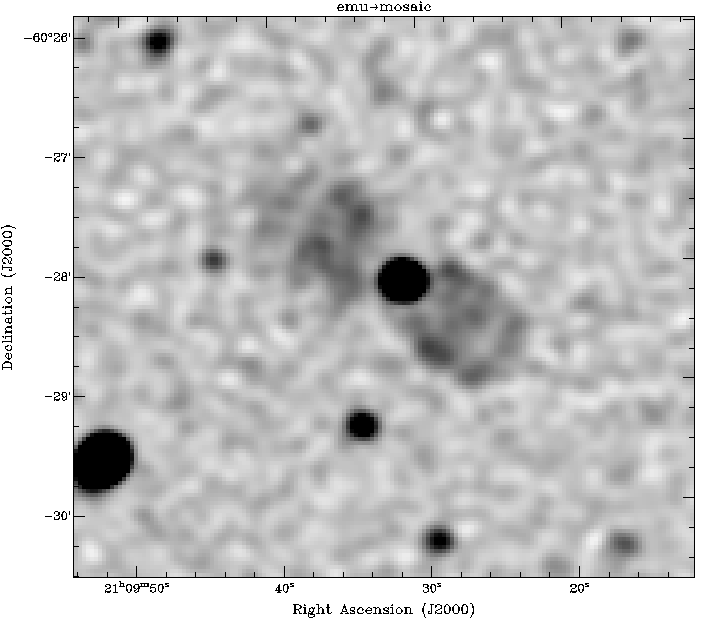}
\caption{EMU-PS image of the edge-brightened double-lobe radio galaxy EMU~PD~J210931.3-602806. \alter{No redshift is available for this source.}}
\label{fig:bipolar}
\end{center}
\end{figure}

Another possibility is that a population of  radio galaxies with edge-brightened lobes, such as the one in Figure~\ref{fig:bipolar}, could provide the parent population. This source has a total extent of 150" and a width of 40", with an edge on the southern lobes that is 30-50\% brighter than the central regions of the lobe. Such extremely faint sources would be detected more easily when viewed end-on, where they would appear ORC-like.  However, such sources would still have  a bright central AGN, which is not seen in all ORCs, and \alter{this model suffers from the same problem as the previous model:} the inferred transverse sizes \alter{of the lobes} of hundreds of kpc 
\alter{is an order of magnitude greater than  is observed in other radio galaxies.} 

ORC-like structures could also appear from the end-on view of bubbles or tori from buoyantly rising old lobes of radio galaxies in a cluster atmosphere   \citep{churazov,randall}. In a later stage of evolution, these could even be re-energized and become visible through collisions with shocks    \citep{Kang} and be responsible for elliptical ring-like radio ``relics'' produced during cluster mergers   \citep{bagchi06,paul11,kale}, if seen edge on. However, these suffer from the problems that (a) such relics do not have the strong circular symmetry of the ORCs, and (b) \alter{as discussed in Section \ref{cluster}}, no known cluster is  associated with the ORCs.

Another possibility is that a radio galaxy can leave behind a blob of plasma, which can then form a vortex ring if it encounters a shock.  Although the initial shape of the ring is likely to be irregular,  reflecting the irregular intersection cross section, the vortex ring is self-propagating and tends to circularize with time   \citep{nolting1,nolting2}. Such a blob/shock encounter was proposed  to explain the large ring at the end of NGC~1265   \citep{pfrommer}.  After the AGN has turned off, the re-energized synchrotron-emitting ring may not show any obvious connection to its parent galaxy. \alter{We consider this to be a possible, although unproven,  mechanism to explain the ORCs.}

\subsection{Cluster halo}
\label{cluster}
Clusters of galaxies often show diffuse radio halos about an arcmin \revb{or more} in diameter   \citep{feretti12, vanweeren}. However, their morphology is typically irregular, and 
punctuated by radio emission from their constituent galaxies, and sometimes with a diffuse relic towards the edge. 
Radio halo brightness profiles typically peak at the centre and decrease radially without any ring-like structures. They are often observed as patchy, but none, to the best of our knowledge, shows the circularly symmetric edge-brightening seen in the ORCs. Furthermore, cluster halo emission is generally accompanied by a cluster of galaxies, and no cluster of galaxies is seen within the ORCs, nor listed in any of the cluster catalogues  covering this area (listed by Manojlovi\'c et al., 2020, in preparation).

We therefore consider it unlikely 
that the ORCs are cluster halos. The possibility that ORCs are related instead to radio ``relics'' seen at the peripheries of merging clusters is discussed in Section \ref{RG2}.

\subsection{Einstein Ring}
Gravitational lensing of background sources can produce arcs of emission. 
If the source, lens, and observer are aligned, then the lensed image can take the form of a so-called Einstein ring, \revb{typically detected in the optical \citep[e.g.][]{belokurov07}}. For example, the radio gravitational (compound) lens PKS\,1830--211, 
consists of a $\sim$1 arcsec diameter radio ring   \citep{jauncey91}.
However, such Einstein rings are rarely more than a few arcsec in diameter. Much larger gravitational lenses are known   \citep{zitrin09} but the lensed image in such cases is irregular, consisting of a number of sub-images, because neither the lensing source nor the background source is sufficiently smooth and axisymmetric to produce a circle. A ring similar to the ORCs could in principle be produced by a lensing cluster of mass $2 \times 10^{15} M_\odot$ at a redshift of 1, but (a) there is no sign of a visible cluster within the rings, and (b) it is unlikely that such a lens would be sufficiently symmetric, and perfectly aligned with the background source, to produce the observed circular symmetry.

\section{Conclusion}

\alter{In this paper, we report a potentially new class of radio-astronomical object, consisting of a circular disc seen only at radio wavelengths, which in some cases is limb-brightened, and sometimes contains a galaxy at its centre. We know they are real, because they have been detected using three different telescopes using different software. 

A careful search of the literature has found no previous report of such circular radio objects, other than the well-known objects, such as supernova remnants and starburst rings, which we have shown above can not explain our observations. We therefore propose that they are a new class of radio source.

Such a discovery is not unexpected when a new telescope such as ASKAP observes the sky in a relatively unexplored part of the observational parameter space \citep{norris17,wtf}. In particular, no previous telescope was able to survey large areas of sky with the low-surface-brightness sensitivity, combined with good spatial resolution, of ASKAP.
}

\alter{In this paper, we have shown that none} of the known types of radio object seems able to explain the ORCs as a class.

For example, if the ORCs are Galactic SNRs, which they strongly resemble, then this implies a population of SNRs in the Galaxy \alter{several orders of magnitude} larger than the currently accepted figure, or else a new class of SNR which has not previously been reported. \alter{If the ORCs are starburst rings, which they also strongly resemble, then we would be able to see the stellar ring counterpart in optical observations.}

\rev{We speculate that} the edge-brightening in some ORCs suggests that this circular image may represent a spherical object, which in turn suggests a spherical wave from some transient event. Several such classes of transient events, capable of producing a spherical shock wave, have recently been discovered, such as fast radio bursts   \citep{bannister}, gamma-ray bursts  \citep{Ayal01}, and neutron star mergers  \citep{Hotokezaka15}. However, because of the large angular size of the ORCs, any such transients would have taken place in the distant past. It also seems unlikely that such an event could cause two ORCs together, as seen in \rev{ORCs 2 and 3.}

It is also possible that the ORCs represent a new category of a known phenomenon, such as the jets of a radio galaxy or blazar when seen end-on, down the ``barrel'' of the jet. Alternatively, they may represent some remnant of a previous outflow from a radio galaxy. However, no existing observations of this phenomenon closely resemble the ORCs in features such as their \alter{striking circularity,} the edge-brightening in most cases, or the absence of a visual blazar or radio galaxy at the centre. 

We also acknowledge the possibility that the ORCs may represent more than one phenomenon, and that they have been discovered simultaneously because they match the spatial frequency characteristics of the ASKAP observations, which occupy a part of the observational parameter space which has hitherto been poorly studied. 
\rev{For example, if ORCs 2 and 3 are explained in terms of an unusual radio galaxy, then   a spherical wave from a transient event may explain ORCs 1 and 4}.

\revb{Finally, we note that \citet{kirillov20} have proposed that an ORC may represent the throat of a wormhole. }

Further work is continuing to investigate the nature of these objects.

\begin{acknowledgements}
We thank Heinz Andernach, Paul Nulsen, Tom Jones, Chris Nolting and many EMU members for valuable comments on a draft of this paper, and Stefan Duchesne for helping with the MWA data. \revb{We thank Robert Becker and an anonymous PASA referee for helpful comments}. Partial support for LR comes from U.S. National Science Foundation grant AST17-14205 to the University of Minnesota. 
The Australian SKA Pathfinder is part of the Australia Telescope National Facility which is managed by CSIRO. Operation of ASKAP is funded by the Australian Government with support from the National Collaborative Research Infrastructure Strategy. ASKAP uses the resources of the Pawsey Supercomputing Centre. Establishment of ASKAP, the Murchison Radio-astronomy Observatory and the Pawsey Supercomputing Centre are initiatives of the Australian Government, with support from the Government of Western Australia and the Science and Industry Endowment Fund. We acknowledge the Wajarri Yamatji people as the traditional owners of the Observatory site.
\end{acknowledgements}

\clearpage
\newpage

\section{Data availability.}

All EMU-PS data (tables and images, and uv data) are available from CASDA on \url{http://hdl.handle.net/102.100.100/164555?index=1} or from \url{https://data.csiro.au/dap/public/casda/casdaSearch.zul} under project code AS101. A zoomable image of the EMU-PS survey is available on \url{http://emu-survey.org}.
A listing of observations is on \url{https://apps.atnf.csiro.au/OMP/index.jsp}.

\linespread{1.25}

\bibliographystyle{pasa-mnras}
\bibliography{wtf}

\end{document}